\documentstyle[aps]{revtex}
\newcommand{\ra}{\rangle}
\newcommand{\la}{\langle}
\begin{document}
\title{Distinguishability and indistinguishability by LOCC
}

\author{Heng Fan}
\address{
Quantum computation and information project,
ERATO, Japan Science and Technology Agency,\\
Daini Hongo White Building 201, Hongo 5-28-3, Bunkyo-ku, 
Tokyo 113-0033, Japan.}
\maketitle
                                                        
\begin{abstract}
We show that a set of linearly independent quantum states
$\{ (U_{m,n}\otimes I)\rho ^{AB}(U_{m,n}^{\dagger }
\otimes I)\} _{m,n=0}^{d-1}$,
where $U_{m,n}$ are generalized Pauli matrices,
cannot be discriminated deterministically or probabilistically
by local operations and classical communications (LOCC). 
On the other hand, any $l$ 
maximally entangled states from this set are
locally distinguishable if $l(l-1)\le 2d$. 
The explicit projecting measurements
are obtained to locally discriminate these states.
As an example, we show that four Werner states are
locally indistinguishable.
\end{abstract}
       
The problem of local distinguishability of quantum
states shared by distant parties
has attracting considerable attentions recently.
A number of interesting and often counterintuitive
results have been obtained. 
It should be clear that orthogonal quantum states
can be distinguished, while non-orthogonal
states can only be distinguished probabilistically
if there are no restrictions to measurements. 
If the quantum states are shared by two distant parties,
say Alice and Bob, and only LOCC is allowed,
the possibility of distinguishing these quantum states
will decrease.
Interestingly,
Walgate $et~al$ showed that any two orthogonal pure states
shared by Alice and Bob can be distinguished by LOCC\cite{WSHV,WH}.
On the other hand, there are a set of orthogonal bipartite
pure product states cannot be distinguished with 
certainty by LOCC\cite{BDFMR,BDSW}.

In this Letter, we will show two main results in the following. 
First, we will show that a set of maximally entangled 
states in the standard form
can be discriminated by local projective measurements and
classical communications. Secondly, using the property of
entanglement breaking channel, we will show a set of quantum states
are locally indistinguishable. 

Let's first introduce some notations.
We consider the dimension of the Hilbert space is $d$ which is prime.  
$U_{m,n}=X^mZ^n, m,n=0,\cdots, d-1$
are generalized Pauli matrices constituting an basis
of unitary operators, and $X|j\ra =|j+1{\rm mod}d\ra ,
Z|j\ra =\omega ^j|j\ra , \omega =e^{2\pi i/d}$, where
$\{ |j\ra \}_{j=0}^d$ is an orthonormal basis.
$|\Phi ^+\ra =\frac {1}{\sqrt{d}}\sum _j|jj\ra $.
$|\Phi _{m,n}\ra =(U_{m,n}\otimes I)|\Phi ^+\ra $ is a
basis of maximally entangled states.

Walgate $et~al$ once showed that two Bell states can be distinguished
by LOCC (their result is for the case of arbitrary two orthogonal
states)\cite{WSHV}. On the other hand 
three Bell states are locally distinguisahble probabilistically,
and 
four Bell states are
locally indistinguishable no matther whether the
protocol is  deterministic or 
probabilistic\cite{GKRSS,GKRSSS,HSSH}. 
One straightforward question is 
what is the maximal set of quantum states which are locally 
distinguishable. In particular, we are interested in the following problem:
suppose $\{ (U_{m,n}\otimes I)|\Psi \ra^{AB}\} _{m,n=0}^{d-1}$
is a complete set of maximally entangled states in
$d\otimes d$ system,
are any
$d$ maximally entangled states from this set locally distinguishable?
This set is the best-known complete set of maximally entangled states. 
It is obvious that if we let $|\Psi \ra^{AB}=|\Phi ^+\ra^{AB}
=\sum _j|jj\ra$,
here we omit a normalized factor, then for arbitrary $n_i$,
$d$ maximally entangled states 
$\{ (X^iZ^{n_i}\otimes I)|\Psi \ra^{AB}\} _{i=0}^{d-1}$ are locally 
distinguishable by simply projecting measurements in the computional
basis on both sides, and subsequently by a classical communication.
For the general case, we do not yet have a complete answer to this question.
However, we can obtain a rather general result:

{\it Theorem: Any $l$ maximally entangled states
from the set  $\{ (U_{m,n}\otimes I)|\Psi \ra^{AB}\} _{m,n=0}^{d-1}$
can be distinguished by LOCC if $l(l-1)/2\le d$.}

For example, if $d=2$, then any two $(l=2)$ Bell states are locally
distinguishable, If $d=3$ or $d=5$, 
then any three maximally entangled states
from this set are locally distinguishable.   
An arbitrary maximally entangled state can always be transformed 
to $|\Phi ^+\ra^{AB}=\sum _j|jj\ra$ by
a local unitary operation on one side (A or B, the difference
between the unitary operators on A and B is a transposition).
So, we just need to prove our claim in   
the case  $|\Psi \ra^{AB}=|\Phi ^+\ra^{AB}$.
Let's suppose these $l$ maximally entangled states take
the form  $\{ (X^{m_i}Z^{n_i}\otimes I)|\Phi ^+ 
\ra^{AB}\} _{i=0}^{l-1}$. 
To locally distinguish these states, we first let
$A$ and $B$ do unitary operations $U$ and $V^t$, respectively,
where $t$ is a transposition.
This operation is equivalent to the transformation
$U(X^{m_i}Z^{n_i})V$ on A side. We next will show that
we can find these unitary operators which can transform
these $l$ maximally entangled states to the
set $\{ (X^{m'_i}Z^{n'_i}\otimes I)|\Phi ^+ \ra^{AB}\} _{i=0}^{l-1}$
where there are no equal $m_i'$.
As we mentioned that this set can be simply distinguished
by LOCC. 
Thus we can prove our previous claim.
We remark that unitary operations $U$ and $V^t$
on A,B sides followed by a projective measurements in the
computional basis is equivalent to projective measurements
on A,B sides in two basis corresponding to $U$ and $V^t$. 

As we analyzed, the problem of local distinguishability
now becomes whether we can find two unitary 
operations $U$ and $V$
which transform $\{ X^{m_i}Z^{n_i}\}_{i=0}^{l-1}$
to the set $\{ X^{m'_i}Z^{n'_i}\}_{i=0}^{l-1}$ in which
no $m'_i$ are equal.
We next will give these 
unitary operations. 
The case of $d=2$ is trivial, with the help of the
Hadamard transformation $H_0$, we can always
discriminate two Bell states by
$Z$ basis measurements on both sides. In the following,
we suppose $d\not= 2$.
We define $d$ unitary operators 
$H_{\alpha }$, $(\alpha =0,1,\cdots ,d-1)$ like this,
the entries of matrices $H_{\alpha }$ take the form
\begin{eqnarray}
\left( H_{\alpha }\right) _{jk}=
\omega ^{-jk}\omega ^{-\alpha s_k},
~~j,k=0,\cdots ,d-1,
\nonumber \\
s_k=k+(k+1)+\cdots +(d-1).  
\label{hadamard}
\end{eqnarray}
By using $H_{\alpha }$, we have the following relations
\begin{eqnarray}
&&H_{\alpha }XH_{\alpha }^{\dagger }=Z^{-1}X^{\alpha },
\nonumber \\
&&H_{\alpha }ZH_{\alpha }^{\dagger }=X.
\end{eqnarray}
Thus $H_{\alpha }$ can transforms $U_{m_i,n_i}$ as follows,
\begin{eqnarray}
H_{\alpha }X^{m_i}Z^{n_i}H_{\alpha }^{\dagger }=
X^{m_i\alpha +n_i}Z^{-m_i}
\label{trans}
\end{eqnarray}
up to a whole phase. 
Given $l$ maximally entangled states
corresponding to $\{ X^{m_i}Z^{n_i}\}_{i=0}^{l-1}$,
we can always transform them to the case where
the powers of $X$ are different
by identity (do nothing) or $H_{\alpha }, \alpha =0,\cdots ,d-1$. 
If not that means for each transformation
always at least two powers of $X$ are equal. 
So we have at least $d+1$ equations altogether. But different combinations
between $l$ elements $\{(m_i,n_i)\}_{i=0}^{l-1}$ is 
$\left( \begin{array}{c}l\\2 \end{array}\right)=l(l-1)/2$ which
is less than or equal to $d$. 
That means two pairs,  for example,
$(m_0,n_0)$ and $(m_1,n_1)$ without loss of generality
will appear twice in two different transformations, say
$\alpha _0$ and $\alpha _1$.
Thus we should have the following relations,
\begin{eqnarray}
\alpha _0m_0+n_0=\alpha _0m_1+n_1, ~({\rm mod}d)
\nonumber \\
\alpha _1m_0+n_0=\alpha _1m_1+n_1.~({\rm mod}d)
\end{eqnarray}
That means $(m_0,n_0)=(m_1,n_1)$ which contradicts with
our assumption that
these $l$ maximally entangled states are orthogonal.
This completes our proof.

We next clarify our proof in the case $d=3$.
Explicitly, the three operators $H_{\alpha }$
take the following form
\begin{eqnarray}
H_0=\left( \begin{array}{ccc}1&1&1 \\
1&\omega ^2&\omega \\
1&\omega &\omega ^2\end{array}\right) ,
\nonumber \\
H_1=\left( \begin{array}{ccc}1&1&\omega \\
1&\omega ^2&\omega ^2\\
1&\omega &1\end{array}\right) ,
\nonumber \\
H_2=\left( \begin{array}{ccc}1&1&\omega ^2 \\
1&\omega ^2&1 \\
1&\omega &\omega \end{array}\right) .
\end{eqnarray}
Given three maximally entangled states corresponding to
$\{X^{m_i}Z^{n_i}\} _{i=0}^2$, if $\{m_i\} _{i=0}^2=\{0,1,2\}$,
it is obvious that they are distinguishable by LOCC.
If $\{n_i\} _{i=0}^2=\{0,1,2\}$, by transformation
$H_0$, they can be distinguished locally. 
The left unsolved cases have the form
$\{(m_0,n_0),(m_0,n_1),(m_2,n_0)\}$ where $n_0\not= n_1, 
m_0\not =m_2$. This form can neither be locally distinguished by
direct measurements in the computional basis nor can be
distinguished by $H_0$ followed by measurements in the computional 
basis. But $H_1$ or $H_2$ will transfer the power of $X$ to
the set $\{0,1,2\}$. If not that means
\begin{eqnarray}
m_0+n_1=m_2+n_0,~{\rm mod}3.
\nonumber \\
2m_0+n_1=2m_2+n_0,~{\rm mod}3.
\end{eqnarray}
Then we know $m_0=m_2, n_0=n_1$ which contradict with
our assumption.
So,  
{\it any 3 maximally entangled states
from the set  $\{ (U_{m,n}\otimes I)|\Psi \ra^{AB}\} _{m,n=0}^{2}$
can be distinguished by LOCC}.

We give some examples to show our local discrimination method.
We have three maximally entangled states 
$|00\ra +\omega |11\ra +\omega ^2|22\ra $,
$|00\ra +\omega ^2|11\ra +\omega |22\ra $ and
$|10\ra +|21\ra +|02\ra $.
corresponding to set $\{ Z,Z^2,X\}$. 
They cannot be discriminated directly by measurements
in computional basis.
By transformation
$H_0$, these three maximally entangled states can be
transformed as $\{X,X^2,Z^2\}$ corresponding to
$|10\ra +|21\ra +|02\ra $,$|20\ra +|01\ra +|12\ra $,
$|00\ra +\omega ^2|11\ra +\omega |22\ra $ which
can be discriminated by projective measurements in
the computional basis followed by a classical communication.
More explicit, if the powers of $Z$ are different in
the given set, we can discriminate them by transformation $H_0$
which is essentially Hadamard transformation.
Suppose the given set is $\{ X,Z,XZ\}$ 
corresponding to $|10\ra +|21\ra +|02\ra $,
$|00\ra +\omega |11\ra +\omega ^2|22\ra $, 
$|10\ra +\omega |21\ra +\omega ^2|02\ra $
which cannot 
be discriminated by transformations $H_0$ and $H_1$.
By transformation $H_2$, we have 
$\{XZ^2,X^2,Z^2\}$ corresponding to
$|10\ra +\omega ^2|21\ra +\omega |02\ra $,
$|20\ra +|01\ra +|12\ra $ and
$|00\ra +\omega ^2|11\ra +\omega |22\ra $ which
can be simply discriminated locally.
The whole procedure is like the following,
A and B do unitary transformations $H_2$ and $H_2^*$ so that
the three maximally entangled states corresponding to
$\{XZ^2,X^2,Z^2\}$. Then A and B do measurements in
the computional basis and subsequently by a classical
communication can discriminate these three maximally
entangled states. 

In general $d$ case, $d$ independent transformations
$H_{\alpha }$ is not enough
to locally distinguish arbitrary $d$ maximally entangled states in
the set  $\{ (X^{m_i}Z^{n_i}\otimes I)|\Phi ^+ 
\ra^{AB}\} _{i=0}^{d-1}$. 
So, we need to find other transformations.
Here we remark that any transformation which changes the power of
$X$ to $jm_i+kn_i$, $j,k=0,\cdots, d-1$
cannot provide new transformations different 
from identity and these $d$ transformations
$H_{\alpha }$ which change the power of $X$ as
$\alpha m_i+n_i$ as shown in (\ref{trans}). 

Combine the result in this Letter for 3-dimension
and the fact that two Bell states can be distinguished locally,
we can generalize our result to $(2\otimes 2)^{\otimes M}
\otimes (3\otimes 3)^{\otimes N}$ case.
Suppose $|\Psi \ra ^{AB}$ is a maximally entangled state in
$2^M3^N\otimes 2^M3^N$ system, then
$2^M3^N$ maximally entangled states from the
set $\{(U_{\vec{m},\vec{n}}\otimes I)|\Psi \ra ^{AB}\}$
can be distinguished by LOCC, where 
$U_{\vec{m},\vec{n}}$ are tensor product of identity and
Pauli matrices
in 2 and 3-dimensions.
Certainly, similar result based on our theorem for general
$\prod \otimes (d_i\otimes d_i)$ case can also be presented.

Horodecki $et~al$ showed that an arbitrary complete set of
orthogonal states of any bipartite system is locally 
indistinguishable if at least one of the vectors is
entangled\cite{HSSH}.
Next we will show the following
result:
An ensemble of linearly independent quantum states
 $\{ \rho _{m,n}^{AB}\} _{m,n=0}^{d-1}$
cannot be discriminated deterministically or probabilistically
by LOCC, where $\rho _{m,n}^{AB}=
(U_{m,n}\otimes I)\rho ^{AB}(U_{m,n}^{\dagger }\otimes I)$. 
We remark that the quantum states of this ensemble
are generally mixed states. 
And this set may 
includes both orthogonal and non-orthogonal quantum states.

We say a quantum channel $\Lambda $ is entanglement breaking
if for all input states, the output states of
the channel $\Lambda \otimes I$ are separable states.
We define a quantum channel $\Lambda ^{AC}$ as follows
\begin{eqnarray}
\Lambda ^{AC}(\rho ^{AC})=
\sum _{mn}U_{m,n}\otimes U_{m,-n}(\rho ^{AC})
U_{m,n}^{\dagger }\otimes U_{m,-n}^{\dagger }.
\label{define}
\end{eqnarray}
Next, we will prove that this quantum channel is entanglement breaking.
To prove that a quantum channel $\Lambda $ is entanglement breaking,
it is enough to show that $\Lambda \otimes I$ maps a maximally
entanglement state into a separable state\cite{CDKL,VDC,HSR}.
Considering that the quantum state
$|\Phi _{0,0}^{AB}\ra \otimes |\Phi _{0,0}^{CD}\ra$
of four systems 
${\cal H}_A\otimes {\cal H}_B\otimes 
{\cal H}_C\otimes {\cal H}_D$ is a maximally entangled state
across $AC:BD$ cut, we should show that
\begin{eqnarray}
&&\Lambda ^{AC}\otimes I^{BD}(\Phi _{0,0}^{AB}\otimes
\Phi _{0,0}^{CD})
\nonumber \\
&=&
\frac {1}{d^2}\sum _{mn}|\Phi _{m,n}^{AB}\ra \la \Phi _{m,n}^{AB}|
\otimes |\Phi _{m,-n}^{CD}\ra \la \Phi _{m,-n}^{CD}|
\end{eqnarray}
is a separable state. 
Actually, we have the
following symmetry,
\begin{eqnarray}
&&\frac {1}{d^2}\sum _{mn}|\Phi _{m,n}^{AB}\ra \la \Phi _{m,n}^{AB}|
\otimes |\Phi _{m,-n}^{CD}\ra \la \Phi _{m,-n}^{CD}|
\nonumber \\
&=&\frac {1}{d^2}\sum _{kl}|\Phi _{k,l}^{AC}\ra \la \Phi _{k,l}^{AC}|
\otimes |\Phi _{k,-l}^{BD}\ra \la \Phi _{k,-l}^{CD}|.
\label{symmetry}
\end{eqnarray}
It is obvious that this is a separable state across $AC:BD$ cut.
Thus we show that $\Lambda ^{AC}$ defined in Eq.(\ref{define})
is an entanglement breaking channel.
Eq.(\ref{symmetry}) can be proved like
the following, we substitute the
following relation,
\begin{eqnarray}
|\Phi _{0,0}^{AB}\ra \otimes |\Phi _{0,0}^{CD}\ra
=\frac {1}{d}\sum _{m,n}|\Phi _{m,n}^{AC}\ra 
\otimes |\Phi _{m,-n}^{BD}\ra .
\end{eqnarray} 
into $\Lambda ^{AC}\otimes I^{BD}
(|\Phi _{0,0}^{AB}\ra \otimes |\Phi _{0,0}^{CD}\ra )$.
With the help of the relation $U_{m,n}U_{k,l}=\omega ^{nk-ml}
U_{k,l}U_{m,n}$, and also we know $|\Phi _{0,0}\ra $ is invariant
under the action of $U_{m,n}\otimes U_{m,-n}$, one can
readily show Eq.(\ref{symmetry}).
We remark that the quantum state (\ref{symmetry}) is
the so-called unlockable bound entangled state in 
$d$-dimension\cite{S}.

Now we are ready for our result of local indistinguishability.
Given the set of linearly independent 
states $\{ \rho _{m,n}^{AB}\} _{m,n=0}^{d-1}$ to 
be discriminated, 
we can construct a quantum state
\begin{eqnarray}
\rho &=&
\frac {1}{d^2}\sum _{mn}\rho _{m,n}^{AB}
\otimes |\Phi _{m,-n}^{CD}\ra \la \Phi _{m,-n}^{CD}|.
\nonumber \\
&=&\Lambda ^{AC}\otimes I^{BD}(\rho ^{AB}\otimes \Phi ^{CD}).  
\end{eqnarray}
Here the maximally entangled states 
$\{ |\Phi _{m,-n}^{CD}\ra \}$ act as detectors. 
Since we know that the quantum channel $\Lambda ^{AC}$ is
entanglement breaking, so this mixed state $\rho $ is 
a separable state across $AC:BD$ cut.
Thus we can show that:{\it A set of linearly independent
quantum states 
 $\{\rho _{m,n}\} _{m,n=0}^{d-1}$ cannot be distinguished
deterministically or probabilistically by LOCC} 
\cite{comment}.
Because if they can be distinguished deterministically
or probabilistically, one could distill non-zero entanglement
by LOCC. This contradicts with the observation that
$\rho $ is a separable state across $AC:BD$ cut.
Note that $\{\rho ^{AB}_{m,n} \}_{m,n=0}^{d-1}$ is in 
$d\otimes d'$ system, and $d,d'$ are not
necessarily the same.
We also should point out that 
if $\rho ^{AB}=|\Psi ^{AB}\ra \la \Psi ^{AB}|$ which 
is a pure state, 
$\{|\Psi _{m,n}\ra \}_{m,n=0}^{d-1}$ are not necessarily orthogonal
to each other, where we denote $|\Psi _{m,n}\ra =U_{m,n}
\otimes I|\Psi \ra $. So, this case is not covered by the 
result in Ref.\cite{GKRSSS,HSSH}. Certainly, distinguishability
of non-orthogonal states is less than that of orthogonal states,
but still they 
can be distinguished probabilistically by global measurements
and for some cases by LOCC\cite{VSPM,CY}.
We will not discuss the case that
$\{|\rho ^{AB}_{m,n}\}_{m,n=0}^{d-1}$ are
linearly dependent.

We next give three examples:

{\it Example 1}:  
According to our result,
an ensemble of states $|\Psi _{0,0}\ra 
=\alpha |00\ra +\beta |11\ra $,
$|\Psi _{0,1}\ra =\alpha |00\ra -\beta |11\ra $,
 $|\Psi _{1,0}\ra =\alpha |10\ra +\beta |01\ra $
and  $|\Psi _{1,1}\ra =\alpha |10\ra -\beta |01\ra $ cannot be distinguished
by LOCC\cite{comment2}. Here we do not consider the special cases such
as $\alpha \beta =0$ which lead to result that the quantum states
of this ensemble are linearly dependent. 
One can find that $|\Psi _{0,0}\ra $ and $|\Psi _{0,1}\ra $
are generally non-orthogonal, while they are orthogonal with
$|\Psi _{1,0}\ra , |\Psi _{1,1}\ra $. So, this ensemble
consists of both orthogonal and non-orthogonal states. And
this case is not studied previously. 
As a special case, we can show that four Bell states
cannot be distinguished by LOCC which has already been pointed out in 
Ref.\cite{GKRSS}.

{\it Example 2}:
We can choose a quantum state in
$|\Psi ^{AB}\ra $ in $2\otimes d'$ system, say let $d'=4$.
For example, let 
$|\Psi ^{AB}\ra =\frac {1}{2}
(|00\ra +|01\ra +|12\ra +|13\ra )$, and 
we have four orthogonal states
$\{(U_{m,n}\otimes I)|\Psi ^{AB}\ra \}_{m,n=0}^1$.
According to our criterion,
they cannot be distinguished by LOCC. Horodecki $et~al$ 
once showed that an arbitrary complete set of orthogonal
states in bipartite system cannot be distinguished
by LOCC if at least one of the states
is entangled, deterministically or probabilistically\cite{HSSH},
For example, four Bell states cannot be distinguished
by LOCC\cite{GKRSSS}. On the other hand, 
three Bell states which is incomplete
can be distinguished probabilistically, and
two Bell states can be distinguished deterministically\cite{WSHV}.
An interesting question is whether there exist 
incomplete sets of orthogonal states which
cannot be distinguished even probabilistically.
Here we present an example to show that there exist
an incomplete set of orthogonal states which cannot
be distinguished by LOCC no matter whether
the protocol is deterministic or probabilistic.
Certainly, we can also give an example of non-orthogonal
states with the same property.
In Ref.\cite{HSSH}, Horodecki $et~al$ presented
an example of incomplete set of
orthogonal states which is indistinguishable by LOCC
deterministically. However, it is still possible that
this set is local indistinguishable probabilistically. 
One may point out that these four states are essentially
four Bell states. It's true. But our conclusion
is not trivial. In general for a bipartite system 
$d\otimes d'$,
there exist $d^2$  orthogonal states which
cannot be distinguished by LOCC \cite{comment3},
even probabilistically.
These $d^2$ orthogonal states are not a complete set if
$d\not= d'$.

{\it Example 3}: Our result is generally for mixed states. 
For qubits case,  
let $\rho ^{AB}=\frac {4p-1}{3}
|\Phi ^+\ra \la \Phi ^+|+\frac {1-p}{3}I$ be the Werner state.
Then we know four different Werner states set
$\{(U_{m,n}\otimes I)\rho ^{AB}(U_{m,n}^{\dagger }\otimes I)
\}_{m,n=0}^1$ are locally indistinguishable, where 
$p\not= 1/4$.

We can generalize
the previous result to states in $2^N\otimes 2^N$ case. 
It is straightforward to show that 
$\{ |\Psi _{m_1,n_1}^{A_1B_1}\ra \otimes 
|\Psi _{m_2,n_2}^{A_2B_2}\ra \otimes \cdots 
\otimes |\Psi _{m_N,n_N}^{A_NB_N}\ra \}_{m_i,n_i=0}^1$ are
indistinguishable by LOCC across 
$A_1\cdots A_N:B_1\cdots B_N$ cut irrespective the
protocol is
deterministic or probabilistic, where
$|\Psi _{m_i,n_i}^{A_iB_i}\rangle =(U_{m_i,n_i}\otimes I)
(\alpha _i|00\ra +\beta _i|11\ra ),\alpha _i\beta _i\not= 0$, and
$m_i,n_i=0,1$. 

Similarly, we can study a more general case
of $\prod _{i=1}^N\otimes (d_i\otimes d_i')$ system.
We define the quantum channel $\Lambda ^{AC}$,
in Hilbert space ${\cal H}_A\otimes {\cal H}_C$, where
${\cal H}_A={\cal H}_{A_1}\otimes \cdots \otimes
{\cal H}_{A_N}$, similarly for ${\cal {H}}_C$.
We can find that the quantum channel defined as
\begin{eqnarray}
\Lambda ^{AC}(\rho ^{AC})=
\sum _{\vec{m}\vec{n}}U_{\vec{m},\vec{n}}\otimes 
U_{\vec{m},-\vec{n}}(\rho ^{AC})
U_{\vec{m},\vec{n}}^{\dagger }\otimes U_{\vec{m},-\vec{n}}^{\dagger }.
\label{define1}
\end{eqnarray}
is an entanglement breaking channel,
where we use the notations $U_{\vec{m},\vec{n}}
=U_{m_1,n_1}\otimes \cdots \otimes U_{m_N,n_N}$.
And thus we can show that for an ensemble of 
linearly independent quantum states
$\{ \rho _{m_1,n_1}^{A_1B_1}\otimes \cdots 
\otimes \rho _{m_N,n_N}^{A_NB_N} \}_{m_i,n_i=0}^{d-1}$,
they can neither be distinguished deterministically nor 
probabilistically. Note that  $A$ side has 
subsystems $A_1\otimes \cdots \otimes A_N$ and collective
measurements are allowed in discrimination. 
But $A$ and $B$ are spatially separated parties
and only classical communication is allowed.

Horodecki $et~al$ also proposed a method to construct a pure 
quantum state by the superposition rather than the mixture\cite{HSSH}.
Then by Jonathan-Plenio criterion \cite{JP} based on
majorization scheme\cite{N,V}, one can check whether
the given quantum states can be distinguished or not if only LOCC
is allowed. Generally, this method relies on some numerical search
which may be complicated. 
Chefles recently 
showed a necessary and sufficient condition for LOCC
unambiguous state discrimination\cite{C}.
In this Letter, we develop
the method of constructing a mixed state 
\cite{TDL,GKRSS,GKRSSS,HSSH},
then by the definition of entanglement breaking channel to
show a family of states are indistinguishable by LOCC,
deterministically or probabilistically.

We show that the quantum channel defined in (\ref{define}) is
an entanglement breaking channel, thus lead to some interesting
results. The method to correspond the entanglement breaking
channel with the indistinguishability by LOCC is a rather powerful
method. Assume that the following quantum channel is 
entanglement breaking
\begin{eqnarray}
\Lambda ^{AC}(\rho ^{AC})
=\sum _iA_i\otimes C_i(\rho ^{AC})A_i^{\dagger }\otimes
C_i^{\dagger }.
\end{eqnarray}
And suppose that the set of quantum states 
$\{ (A_i\otimes I)\rho ^{AB}(A_i^{\dagger }\otimes I) \}$ with
normalization to be distinguished   
are linearly independent, and we assume that  not all detectors
$\{ (C_i\otimes I)\rho ^{CD}(C_i^{\dagger }\otimes I) \}$
are separable states. Here the detector $\rho ^{CD}$ should be 
an entangled state but not necessarily a maximally entangled state. 
With the input state of the channel taking 
$\rho ^{AB}\otimes \rho ^{CD}$, we know the output state is a separable state
across $AC:BD$ cut. Thus we know this 
set of states
$\{ (A_i\otimes I)\rho ^{AB}(A_i^{\dagger }\otimes I) \}$ 
cannot be distinguished by
LOCC\cite{comment2}.
 
In summary, we proposed a family of unitary transformations
$\{ H_{\alpha }\}_{\alpha =0}^{d-1}$ in (\ref{hadamard}).
By projective measurements corresponding to
these transformations, we can locally discriminate any
$l$ maximally entangled states choosed from the set
$\{ (U_{m,n}\otimes I)|\Psi \ra^{AB}\} _{m,n=0}^{d-1}$ 
if $l(l-1)\le 2d$. And from the property of entanglement
breaking, we show that a family of quantum states
are indistinguishable by LOCC.
 
{\it Acknowlegements}: The author would like to thank
K.Matsumoto, M.Plenio,T.Shimono, and X.B.Wang   
for useful discussions and suggestions.


\begin{thebibliography}{99}
\bibitem{comment}In this Letter, we only consider the case that one copy
of the quantum state is provided.
\bibitem{comment2}We say that the quantum states cannot be distinguished
by LOCC means that they cannot be distinguished by LOCC deterministically
or probabilistically.
\bibitem{comment3}Without lose of generality, we assume $d'\ge d$.
\end{thebibliography}
\end{document}